\documentstyle[twoside,fleqn,espcrc2]{article}

\newcommand{\AmS}{{\protect\the\textfont2
  A\kern-.1667em\lower.5ex\hbox{M}\kern-.125emS}}

\title{Deriving exact results for Ising-like models from the cluster
variation method}

\author{Alessandro Pelizzola\address{Istituto Nazionale per la Fisica
           della Materia and 
           Dipartimento di Fisica del Politecnico di Torino,        
           c. Duca degli Abruzzi 24, 10129 Torino, Italy}%
        \thanks{E-mail: alex@athena.polito.it}}
       
\begin{document}

\begin{abstract}
The cluster variation method (CVM) is an approximation
technique which generalizes the mean field approximation and has been
widely applied in the last decades, mainly for finding accurate phase
diagrams of Ising-like lattice models. Here we discuss in which cases
the CVM can yield exact results, considering: (i) one-dimensional
systems and strips (in which case the method reduces to the transfer
matrix method), (ii) tree-like lattices and (iii) the so-called
disorder points of euclidean lattice models with competitive
interactions in more than one dimension.
\end{abstract}

\maketitle

The cluster variation method (CVM) is a hierarchy of approximation
techniques for discrete (Ising-like) classical lattice models, which has been
invented by Kikuchi \cite{kik1}. In its modern formulation \cite{an}
the CVM is based on the variational principle of equilibrium
statistical mechanics, which says that the free energy $F$ of a
model defined on the lattice $\Lambda$ is given by
\begin{equation}
F = {\rm min} \ F[\rho_\Lambda] = {\rm min} \ 
{\rm Tr}(\rho_\Lambda H + \rho_\Lambda \ln \rho_\Lambda),
\end{equation}
where $H$ is the hamiltonian of the model, $\beta = 1$ for simplicity,
and the density matrix $\rho_\Lambda$ must be properly normalized:
${\rm Tr}(\rho_\Lambda) = 1$. 

As a first step one usually introduces the cluster density matrices
and the cluster entropies 
\begin{equation}
\rho_\alpha = {\rm Tr}_{\Lambda \setminus \alpha} \rho_\Lambda  \qquad
S_\alpha = - {\rm Tr} (\rho_\alpha \ln \rho_\alpha),
\end{equation}
where $\alpha$ is a cluster of $n_\alpha$ sites and ${\rm Tr}_{\Lambda
\setminus \alpha}$ denotes a summation over all degrees of freedom
except those belonging to the cluster $\alpha$. One then introduces
the cumulant expansion of the cluster entropies
\begin{equation}
S_\alpha = \sum_{\beta \subseteq \alpha} \tilde S_\beta
\quad \Leftrightarrow \quad \tilde S_\beta = \sum_{\alpha \subseteq \beta}
(-1)^{n_\alpha - n_\beta} S_\alpha,
\end{equation}
in terms of which the variational free energy can be rewritten as
\begin{equation}
F[\rho_\Lambda] = {\rm Tr}(\rho_\Lambda H) - \sum_{\beta
\subseteq \Lambda} \tilde S_\beta.
\end{equation}

The above steps are all exact and the approximation defining the CVM
comes in when one truncates the cumulant expansion of the
entropy. The sum of the cumulants of the cluster entropies is
restricted to a given set $M$ of clusters, which in most cases can be
thought of as a set of maximal clusters and all their subclusters. If
the model under consideration has only short range interactions and the
maximal clusters are sufficiently large the hamiltonian can be
decomposed into a sum of cluster contributions and the approximate
variational free energy takes the form
\begin{eqnarray}
F[\{\rho_\alpha, \alpha \in M\}] &\simeq&
\sum_{\alpha \in M} \left[ {\rm Tr}(\rho_\alpha H_\alpha) - \tilde
S_\alpha \right] \\
&=& \sum_{\alpha \in M} \left[ {\rm Tr}(\rho_\alpha H_\alpha) -
a_\alpha S_\alpha \right] \nonumber,
\end{eqnarray}
where the coefficients $a_\alpha$ can be easily obtained from the set
of linear equations
\begin{equation}
\sum_{\beta \subseteq \alpha \in M} a_\alpha = 1, \qquad \forall
\beta \in M
\label{CVMcoefs}
\end{equation}
and the cluster density matrices must satisfy the following
conditions which express normalization
\begin{equation}
{\rm Tr} \rho_\alpha = 1, \qquad \forall \alpha \in M 
\end{equation}
and compatibility
\begin{equation}
\rho_\alpha = {\rm Tr}_{\beta \setminus \alpha} \rho_\beta, \qquad \forall
\alpha \subset \beta \in M.
\end{equation}

Having introduced an approximation it is worth asking whether there
are special cases in which it turns out to be exact. The simplest
example is that of a system defined on a lattice $\Lambda$ which can
be regarded as the union of two clusters $\Lambda = A \cup B$, such that,
denoting by $K = A \cap B$ their intersection,
there is no interaction between $A^\prime = A \setminus K$
and $B^\prime = B \setminus K$. In this case the hamiltonian has the
general form $H = H_A(\underline\sigma_A) + H_B(\underline\sigma_B)$ and
it is easy to check that the density matrix can be written as
$\rho_\Lambda = \displaystyle\frac{\rho_A \rho_B}{\rho_K}$, which in
turn implies the decomposition $S_\Lambda = S_A + S_B - S_K$ for the
entropy. The CVM approximation which one obtains with the set of
clusters $M = \{A,B,K\}$ leads to the same decomposition of the
entropy (eq.\ \ref{CVMcoefs} yields $a_A = a_B = 1$, $a_K = -1$) and
is therefore exact. It can be verified that this argument can be
easily generalized (to several clusters sharing a common intersection)
and/or iterated (to a chain of clusters $A,B,C, \ldots$).

The above argument could be used to explain the well-known fact that
CVM approximations are exact for Bethe and cactus lattices (that is,
interior of Cayley and Husimi trees, respectively), which are made of
links (respectively plaquettes), sharing common sites with no loops
(respectively no loops larger than the elementary plaquette). However
we shall leave apart tree-like lattices and turn our attention to
euclidean ones, considering first one-dimensional systems (strips) and
then the disorder points of higher dimensional models.

Since it is known that the Bethe-Peierls approximation (which is the
lowest level CVM approximation for a model with nearest-neighbour
interactions only, obtained by taking $M = \{links,sites\}$) is exact
for a one-dimensional chain, one might wonder whether there is a CVM
approximation which is exact for a strip of finite width. The answer
is affirmative and it is interesting to note that one recovers the
transfer matrix formalism. Consider a strip of width $N$ and (finite,
for the moment) length
$L$ and let the hamiltonian contain only
translation-invariant NN interactions, with open boundary
conditions. In the above scheme, this is an example of a chain of
intersecting clusters and we can guess that a CVM approximation with
the $N \times 2$ ladders as maximal clusters should be exact. Denoting
by II such clusters and by I their $N \times 1$ intersections we set
$M = \{{\rm II},{\rm I}\}$ (no other subclusters enter the cumulant
expansion) and in the thermodynamic limit $L \to \infty$, assuming that translational
invariance is recovered, we get the variational principle
\begin{eqnarray}
f &=& \lim_{L \to \infty}\frac{F}{L} \nonumber \\
&=& {\rm min} \ {\rm Tr} \left(
\rho_{\rm II} H_{\rm II} +  
\rho_{\rm II} \ln \rho_{\rm II} - \rho_{\rm I} \ln \rho_{\rm I}
\right).
\end{eqnarray}
Denoting by $\underline\sigma$ and $\underline\sigma^\prime$ the two
sets of degrees of freedom of the two I subclusters of a II cluster we
can solve for $\rho_{\rm II}$ and recover the transfer matrix
formalism in the form
\begin{equation}
f = - \ln {\rm max} 
\sum_{\underline\sigma,\underline\sigma^\prime} 
\rho_{\rm I}^{1/2}(\underline\sigma) {\rm e}^{-
H_{\rm II}(\underline\sigma,\underline\sigma^\prime)}
\rho_{\rm I}^{1/2}(\underline\sigma^\prime) 
\end{equation}
with the normalization constraint
$\displaystyle\sum_{\underline\sigma} \rho_{\rm I}(\underline\sigma) = 1$.
It is interesting to note that the CVM comes with a natural fixed point
algorithm \cite{kik2} for finding the local miima of the free energy,
which in this case reduces to the power method for finding the largest
eigenvalue of the transfer matrix. 

The last (and perhaps the most interesting) case we want to consider
is that of disorder points. As an example we consider the square
lattice Ising model with competitive interactions, with hamiltonian
\begin{eqnarray}
H = && - K_1 \sum_{\langle i j \rangle} \sigma_i \sigma_j 
- K_2 \sum_{\langle \langle i j \rangle \rangle} \sigma_i \sigma_j
\nonumber \\
&& - K_4 \sum_{[ijkl]} \sigma_i \sigma_j \sigma_k \sigma_l,
\end{eqnarray}
where $K_1 > 0$ is the NN coupling, $K_2 < 0$ the next nearest
neighbour coupling and $K_4$ the plaquette coupling. It is known
\cite{disorder} that in the disordered phase of this model there is an
integrable subspace given by
\begin{equation}
\cosh (2 K_1) = \frac{{\rm e}^{2K_4} \cosh(4K_2)+
{\rm e}^{-2K_2}} 
{{\rm e}^{2K_2}+{\rm e}^{2K_4}},
\end{equation}
where the free energy density is given by
\begin{equation}
f = - \ln \left[ \exp(-K_4)+\exp(K_4 - 2K_2) \right].
\label{free}
\end{equation}
In this subspace the $R$ matrix has an eigenvector which is a pure tensorial
product and the eigenvector of the transfer matrix corresponding to
the largest eigenvalue is also a pure tensorial product
\cite{disorder}, and hence the 
density matrix and the two-site correlations are factorized. Because
of this factorization (and the corresponding decomposition of the
entropy) one can expect that the model can be solved exactly by the
CVM and indeed this is the case. It is enough to choose $M$ = \{plaquettes
and their subclusters\} (of course larger maximal clusters work as
well) to verify eq.\ \ref{free} and also to calculate the two-site
correlation functions $\Gamma(x,y) = \langle \sigma(x_0,y_0)
\sigma(x_0+x,y_0+y) \rangle$:
\begin{eqnarray}
\Gamma(x,y) &=& g^{|x|+|y|}, \nonumber \\
g &=& \frac{\exp(-4K_2) - \cosh(2 K_1)}{\sinh(2K_1)}
\end{eqnarray}
as well as many-site correlation functions like the plaquette
correlation function $q = \langle \sigma_i
\sigma_j \sigma_k \sigma_l \rangle_{\Box}$ 
\begin{equation}
q = \frac{{\rm e}^{4K_4}\left(1-{\rm e}^{8K_2}\right) +
4{\rm e}^{2K_2}\left({\rm e}^{2K_4}-{\rm e}^{2K_2}\right)} 
{{\rm e}^{4K_4}\left(1-{\rm e}^{8K_2}\right) +
4{\rm e}^{2K_2}\left({\rm e}^{2K_4}+{\rm e}^{2K_2}\right)}.
\end{equation}
Details, and generalizations to other models, will be reported
elsewhere \cite{prep}.


\begin{thebibliography}{9}
\bibitem{kik1} R. Kikuchi, Phys. Rev. {\bf 81} (1951) 988.
\bibitem{an} G. An, J. Stat. Phys. {\bf 52} (1988) 727;
             T. Morita, J. Stat. Phys. {\bf 59} (1990) 819.
\bibitem{kik2} R. Kikuchi, J. Chem. Phys. {\bf 60} (1974)
1071; J. Chem. Phys. {\bf 65} (1976) 4545.
\bibitem{disorder} see e.g.\ I.G. Enting, J. Phys. {\bf C10}
(1977) 1379; R.J. Baxter, J. Phys. {\bf A17} (1984) L911;
H. Meyer et al, Phys. Rev. {\bf E55} (1997) 5380 
and refs.\ therein.
\bibitem{prep} A. Pelizzola, in preparation.
\end{thebibliography}
\end{document}